\documentclass[aps,twocolumn,prl,superscriptaddress]{revtex4-2}
\usepackage[tbtags,fleqn]{amsmath}
\usepackage{xcolor}

\usepackage{amssymb}
\usepackage{amsfonts}
\usepackage{mathtools}
\usepackage[colorlinks=true,citecolor=blue,urlcolor=blue]{hyperref}
\hypersetup{colorlinks,linkcolor=blue,filecolor=blue,urlcolor=blue,citecolor=blue}
\usepackage{multirow}
\usepackage{subfigure}
\usepackage{cases}
\usepackage{xr}

\usepackage{appendix}

\usepackage{times}

\begin{document}
	
	\title{Coherent–Dissipative Crossover of Quantum Synchronization}
	
	\author{Li Liu}
	\affiliation{Key Laboratory of Low-Dimension Quantum Structures and Quantum Control of Ministry of Education, Synergetic Innovation Center for Quantum Effects and Applications, Department of Physics, Hunan Normal University, Changsha 410081, China}
    \affiliation{Hunan Research Center of the Basic Discipline for Quantum Effects and Quantum Technologies, Hunan Normal University, Changsha 410081, China}
	\affiliation{Institute of Interdisciplinary Studies, Hunan Normal University, Changsha 410081, China}

 \author{Niu Wang}
	\affiliation{Key Laboratory of Low-Dimension Quantum Structures and Quantum Control of Ministry of Education, Synergetic Innovation Center for Quantum Effects and Applications, Department of Physics, Hunan Normal University, Changsha 410081, China}
    \affiliation{Hunan Research Center of the Basic Discipline for Quantum Effects and Quantum Technologies, Hunan Normal University, Changsha 410081, China}
	\affiliation{Institute of Interdisciplinary Studies, Hunan Normal University, Changsha 410081, China}
 
    \author{Xun-Wei Xu}
	\affiliation{Key Laboratory of Low-Dimension Quantum Structures and Quantum Control of Ministry of Education, Synergetic Innovation Center for Quantum Effects and Applications, Department of Physics, Hunan Normal University, Changsha 410081, China}
    \affiliation{Hunan Research Center of the Basic Discipline for Quantum Effects and Quantum Technologies, Hunan Normal University, Changsha 410081, China}
	\affiliation{Institute of Interdisciplinary Studies, Hunan Normal University, Changsha 410081, China}
    
\author{Jie-Qiao Liao}
	\affiliation{Key Laboratory of Low-Dimension Quantum Structures and Quantum Control of Ministry of Education, Synergetic Innovation Center for Quantum Effects and Applications, Department of Physics, Hunan Normal University, Changsha 410081, China}
    \affiliation{Hunan Research Center of the Basic Discipline for Quantum Effects and Quantum Technologies, Hunan Normal University, Changsha 410081, China}
	\affiliation{Institute of Interdisciplinary Studies, Hunan Normal University, Changsha 410081, China}

      \author{Hui Jing}
	\affiliation{Key Laboratory of Low-Dimension Quantum Structures and Quantum Control of Ministry of Education, Synergetic Innovation Center for Quantum Effects and Applications, Department of Physics, Hunan Normal University, Changsha 410081, China}
    \affiliation{Hunan Research Center of the Basic Discipline for Quantum Effects and Quantum Technologies, Hunan Normal University, Changsha 410081, China}
	\affiliation{Institute of Interdisciplinary Studies, Hunan Normal University, Changsha 410081, China}

    \author{Le-Man Kuang}
	\affiliation{Key Laboratory of Low-Dimension Quantum Structures and Quantum Control of Ministry of Education, Synergetic Innovation Center for Quantum Effects and Applications, Department of Physics, Hunan Normal University, Changsha 410081, China}
    \affiliation{Hunan Research Center of the Basic Discipline for Quantum Effects and Quantum Technologies, Hunan Normal University, Changsha 410081, China}
	\affiliation{Institute of Interdisciplinary Studies, Hunan Normal University, Changsha 410081, China} 
	
	\author{Changliang Ren}\thanks{Corresponding author: renchangliang@hunnu.edu.cn}
	\affiliation{Key Laboratory of Low-Dimension Quantum Structures and Quantum Control of Ministry of Education, Synergetic Innovation Center for Quantum Effects and Applications, Department of Physics, Hunan Normal University, Changsha 410081, China}
    \affiliation{Hunan Research Center of the Basic Discipline for Quantum Effects and Quantum Technologies, Hunan Normal University, Changsha 410081, China}
	\affiliation{Institute of Interdisciplinary Studies, Hunan Normal University, Changsha 410081, China}
	
	\begin{abstract}
Quantum synchronization in open systems can arise from both coherent interactions and collective dissipation, yet their interplay remains poorly understood. Here we develop a general framework for two quantum oscillators coupled through a fast mediator. Adiabatic elimination reveals two distinct synchronization channels, coherent exchange and collective dissipation, and yields an analytical description connecting dissipation- and coherence-dominated regimes. We show that synchronization is governed by the joint balance of coherent coupling, dissipation, and reservoir-induced fluctuations, rather than by coherent coupling alone. Continuous monitoring of the shared output further converts reservoir information into a resource, substantially enhancing conditional synchronization, with the strongest gain near the coherent--dissipative crossover. These results establish a general mechanism for coherent—dissipative quantum synchronization and its measurement-enabled control in open quantum systems.
	\end{abstract}

	
	\maketitle

	\textit{Introduction-} \textcolor{black}{Synchronization is a ubiquitous emergent behavior in nonlinear dynamical systems, describing the spontaneous formation of temporal coherence among coupled oscillators} \cite{Pikovsky2001,Dorfler2014,RevModPhys.77.137,Arenas2008}. Its quantum extension has opened a new avenue for studying collective nonequilibrium dynamics in open quantum systems \cite{Mari2013,Aifer2024,Bittner2025,PhysRevA.99.033818,PhysRevLett.112.094102,PhysRevLett.121.063601,Giorgi2019,solanki2026quantumsynchronization,schmolke2026synchronizationquantumregime}. Unlike classical synchronization, quantum synchronization is shaped by the interplay of coherent interactions \cite{Kato2021,Mari2013,PhysRevLett.111.234101,PhysRevLett.112.094102,Yang2025}, engineered dissipation \cite{w1bm-wjl4,PhysRevLett.129.250601,yrfl-y3z8}, quantum fluctuations \cite{Hohe2025,Tao2025,PhysRevLett.120.163601,PhysRevLett.117.073601}, and measurement backaction \cite{Kato2021,PhysRevA.111.012202,PhysRevE.108.024204,c5wj-4s73}. These ingredients can generate qualitatively distinct routes to synchronization, as demonstrated across a range of platforms, including cold atoms \cite{PhysRevLett.125.013601,Cabot2019}, nuclear spins \cite{PhysRevA.105.062206}, trapped ions \cite{w1bm-wjl4}, and quantum simulators \cite{PhysRevResearch.2.023026}. Despite this progress, the roles of coherent and dissipative dynamics in establishing quantum synchronization remain intertwined. In particular, whether they represent distinct synchronization mechanisms and how their competition drives a crossover between them remain open fundamental questions in nonequilibrium quantum dynamics.

    Cavity magnonics offers a particularly versatile setting for resolving this question \cite{Harder2021,Wang2020}. In yttrium iron garnet (YIG), magnons couple coherently to microwave cavity modes \cite{PhysRevLett.111.127003,PhysRevLett.113.083603,PhysRevLett.113.156401}, while cavity loss provides a tunable dissipative reservoir \cite{PhysRevLett.121.137203,Wang2020}.\textcolor{black}{A single cavity mode can therefore mediate both coherent exchange and collective dissipative coupling between spatially separated magnons, allowing these two distinct synchronization channels to coexist and be compared within a unified framework \cite{qygs-3kfv,kvrx-8rjf}.}
    This unique architecture provides a controlled setting in which their distinct roles and interplay can be quantitatively characterized.
Despite this advantage, the underlying dynamics remain obscured by the coupled cavity–magnon degrees of freedom \cite{PhysRevLett.121.203601,Zare_Rameshti_2022,Wang2020,Harder2021}, \textcolor{black}{where the relative roles of coherent and dissipative processes in cavity-mediated quantum synchronization, as well as the dominant mechanism, still lack a systematic characterization \cite{7hhn-zypy,Ghildiyal2026,Yang2025}.}In particular, even when the cavity evolves much faster than the magnons, it remains unclear how to systematically eliminate the cavity degrees of freedom and identify the resulting coherent and dissipative synchronization channels, let alone determine how their competition reshapes the synchronization dynamics. A unified effective description of this coherent–dissipative crossover is therefore still lacking.

In this letter, we develop an effective theory of two magnons mediated by a fast cavity and uncover two fundamentally distinct synchronization mechanisms arising from collective dissipation and coherent exchange. By adiabatically eliminating the cavity degree of freedom, we disentangle these two channels and reveal a coherent–dissipative crossover from dissipation-dominated to exchange-dominated synchronization, with their competition determining the emergent synchronization dynamics. This unified picture establishes how coherence and dissipation play distinct yet competing roles in quantum synchronization, providing a direct means to identify and control the underlying synchronization mechanism in open quantum systems.

\textit{Effective quantum dynamics and synchronization.—} We consider an open quantum system comprising two bosonic modes mediated by a single-mode quantum cavity. The model is generic and captures the coherent and dissipative coupling of two quantum oscillators through a shared quantum environment, providing a unified description applicable to a broad range of physical platforms. As a concrete realization, we consider two yttrium iron garnet (YIG) spheres coupled to a microwave cavity and analyze the system in experimentally accessible parameter regimes \cite{PhysRevLett.111.127003,PhysRevLett.113.083603,PhysRevLett.113.156401,Harder2021,PhysRevB.104.064426}, establishing a realistic platform for observing the predicted quantum dynamics.

\begin{figure}[t]
\centering
\includegraphics[width=\columnwidth]{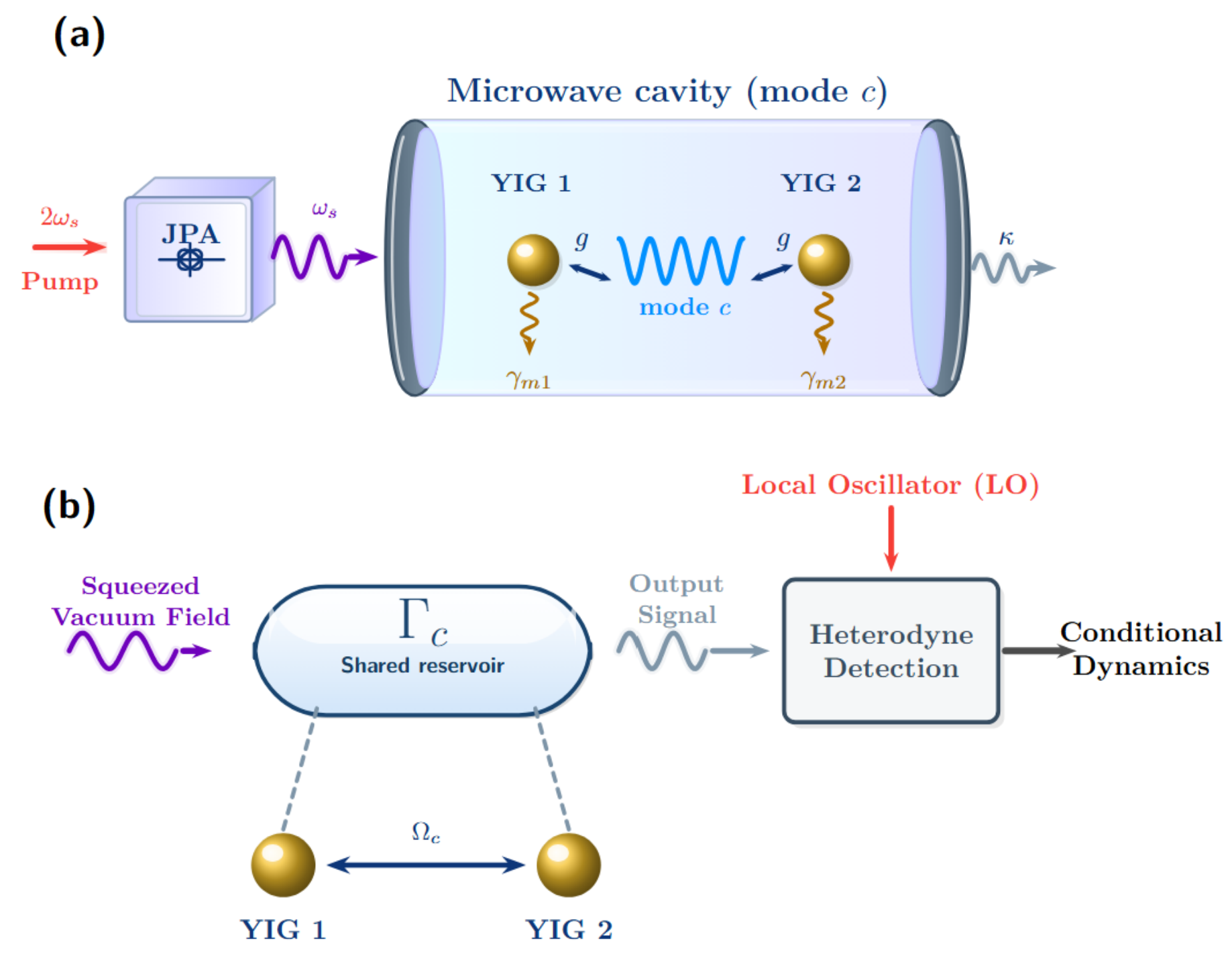}
\caption{Physical configuration of the cavity-magnonic system and schematic of the effective conditional dynamics. (a) Schematic of the full physical setup, where a squeezed vacuum field is injected into the microwave cavity coupled to two YIG magnons. (b) Effective description obtained after adiabatic elimination of the cavity mode, featuring collective dissipation ($\Gamma_c$), coherent magnon exchange ($\Omega_c$), and continuous heterodyne monitoring of the cavity output for conditional state estimation.}
\label{fig:physical_model}
\end{figure}

As illustrated in Fig.~\ref{fig:physical_model}(a), each YIG sphere supports a macroscopic collective spin excitation that, \textcolor{black}{at low excitation densities}, can be mapped onto a bosonic magnon mode $\hat m_{1,2}$ via the Holstein–Primakoff transformation\cite{PhysRevLett.111.127003,PhysRevLett.113.083603,PhysRevLett.113.156401}. The two magnon modes have negligible direct dipolar coupling and instead interact indirectly through their mutual coupling to a single microwave cavity mode $\hat c$\cite{PhysRevB.104.064426}. The cavity is driven by broadband squeezed vacuum generated by a Josephson parametric amplifier, thereby injecting nonclassical quantum fluctuations into the magnon system\cite{7hhn-zypy,PhysRevA.99.021801}. The two magnons are taken to couple symmetrically to the cavity with strength $g$, which is enhanced by the collective spin of the YIG spheres \cite{PhysRevLett.113.156401}.

In a frame rotating at the central frequency $\omega_s$ of the squeezed field, the system Hamiltonian reads
\begin{equation}
\hat H=
\sum_{k=1}^{2}\Delta_{m_k}\hat m_k^\dagger\hat m_k
+\Delta_c\hat c^\dagger\hat c
+ig\left[(\hat m_1^\dagger+\hat m_2^\dagger)\hat c
-\hat c^\dagger(\hat m_1+\hat m_2)\right],
\label{app:Completely Model}
\end{equation}
where $\Delta_{m_k}=\omega_{m_k}-\omega_s$ and $\Delta_c=\omega_c-\omega_s$. Including Markovian coupling of the cavity and magnon modes to their respective environments, the full system obeys the Lindblad master equation
\begin{equation}
\dot\rho=-i[\hat H,\rho]
+\kappa\mathcal D[\hat c]\rho
+\sum_k\gamma_{m_k}\mathcal D[\hat m_k]\rho,
\end{equation}
where $\kappa$ denotes the total decay rate of the microwave cavity and $\gamma_{m_k}$ is the local decay rate of the $k$th magnon mode.

When the dynamical scale introduced by cavity decay and detuning, $|\lambda_c|=\sqrt{\kappa^2/4+\Delta_c^2}$, is much larger than the magnon evolution scale, i.e., $\lambda_{\rm ad}=|\lambda_c|/g\gg1$, the cavity can be treated as a fast variable and adiabatically eliminated. \textcolor{black}{Similar cavity-mode adiabatic elimination has been
employed to engineer effective dissipative dynamics in quantum
self-oscillators\cite{PhysRevLett.118.243602}.}The cavity Langevin equation then gives
\begin{equation}
\hat c\simeq
-\frac{g(\hat m_1+\hat m_2)}
{i\Delta_c+\kappa/2}
+\frac{\sqrt{\kappa}\hat c_{\rm in}}
{i\Delta_c+\kappa/2}.
\end{equation}
Substituting this expression into Eq.~(\ref{app:Completely Model}), the effective two-magnon dynamics are described by the Hamiltonian
$H_{\mathrm{eff}} = \sum_{k=1}^{2} \tilde{\Delta}_{m_k} \hat{m}_k^\dagger \hat{m}_k-\Omega_c (\hat{m}_1^\dagger\hat{m}_2  + \hat{m}_2^\dagger\hat{m}_1)$,
with the corresponding effective Lindblad master equation
$\dot{\rho} = -i[H_{\mathrm{eff}}, \rho]+
\sum_{k=1}^{2} \gamma_{m_k} \mathcal{D}[\hat{m}_k]\rho
+ 2\Gamma_c \mathcal{D}[\hat{m}_1 + \hat{m}_2]\rho$.
Here, $\Gamma_c = \frac{g^2 \kappa/2}{\Delta_c^2 + \kappa^2/4}$ and
$\Omega_c = \frac{g^2 \Delta_c}{\Delta_c^2 + \kappa^2/4}$ denote the cavity-mediated shared dissipative and coherent exchange couplings, respectively, while $\tilde{\Delta}{m_k} = \Delta{m_k} - \Omega_c$ denotes the effective detuning.\textcolor{black}{The appearance of the collective dissipative term
$\mathcal{D}[\hat{m}_1+\hat{m}_2]$ after eliminating the cavity mode
indicates that the two magnons are coupled through a common effective
dissipative channel. Such a collective dissipative structure has also
been considered in studies of quantum synchronization blockade
~\cite{9y9w-lw92}. In the present system, both the shared dissipative
channel and the coherent exchange interaction arise naturally from
the adiabatic elimination of the same cavity mode.}

\begin{figure*}[t]
\includegraphics[width=\textwidth]{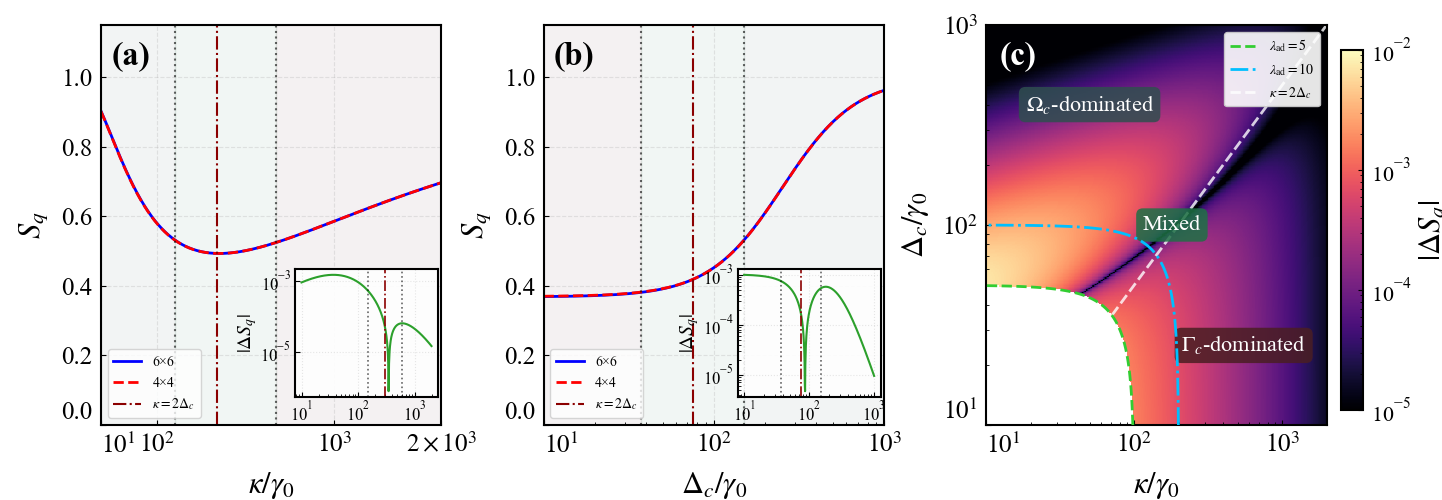}
\caption{Quantitative validation of the adiabatic elimination. (a) Synchronization measure $S_q$ as a function of $\kappa/\gamma_0$ for fixed $\Delta_c/\gamma_0=150$; (b) $S_q$ as a function of $\Delta_c/\gamma_0$ for fixed $\kappa/\gamma_0=150$; (c) absolute deviation $\Delta S_q$ between the full $6\times6$ model and the reduced $4\times4$ model obtained after adiabatic elimination in the $(\kappa,\Delta_c)$ parameter plane. The contours $\lambda_{\mathrm{ad}}=5$ and $\lambda_{\mathrm{ad}}=10$ are included as references for the degree of cavity timescale separation. The blue solid and red dashed curves denote the results from the full $6\times6$ and adiabatically eliminated $4\times4$ models, respectively. Over the parameter range considered, the two models yield closely consistent values of $S_q$, while the deviation $\Delta S_q$ remains small in the fast-cavity regime characterized by increasing $\lambda_{\mathrm{ad}}$, confirming the validity of the adiabatic elimination in the relevant parameter regime.}
\label{fig:adiabatic_validation}
\end{figure*}

The above master equation further yields the Heisenberg–Langevin equations for the magnon modes,
\begin{equation}
\begin{aligned}
\dot{\hat m}_k=&-
(\frac{\gamma{m_k}}2+\Gamma_c+i\tilde\Delta_{m_k})
\hat m_k
-(\Gamma_c-i\Omega_c)\hat m_{k'}
+\eta\hat c_{\rm in}\\
&+\sqrt{\gamma_{m_k}}\hat m_{\rm in,k}\nonumber
\end{aligned}
\end{equation}
where $k'\neq k$ and $\eta = \frac{g\sqrt{\kappa}}{i\Delta_c + \kappa/2}$ characterizes the transfer of cavity input noise to the magnons, with $|\eta|^2 = 2\Gamma_c$. This result reveals a distinct correlation mechanism induced by the shared quantum environment: after adiabatic elimination, the magnons not only acquire cavity-mediated effective interactions but are also driven collectively by the same squeezed input-noise channel.

To characterize the quantum synchronization of the two magnon modes, we introduce the quadrature fluctuations $\delta\hat X_k=(\delta\hat m_k+\delta\hat m_k^\dagger)/\sqrt2$ and $\delta\hat Y_k=(\delta\hat m_k-\delta\hat m_k^\dagger)/(i\sqrt2)$, and define the relative-mode quadratures $\delta\hat X_-=(\delta\hat X_1-\delta\hat X_2)/\sqrt2$ and $\delta\hat Y_-=(\delta\hat Y_1-\delta\hat Y_2)/\sqrt2$. The quantum synchronization measure is defined as \cite{Mari2013,7hhn-zypy,Lai2025} $S_q=1/\Delta_{\mathrm{sync}}^2$, where $\Delta_{\mathrm{sync}}^2=\langle\delta\hat X_-^2\rangle+\langle\delta\hat Y_-^2\rangle$. After eliminating the cavity, the magnons are correlated not only through the effective coherent and dissipative couplings $\Omega_c$ and $\Gamma_c$, but also through the common input-noise term $\eta\hat c_{\rm in}$, which gives rise to nonzero off-diagonal blocks in the diffusion matrix. These noise correlations are determined by the squeezed input field, characterized by the squeezing parameter $r$, thermal photon number $N_a$, and magnon thermal occupation $\mathrm{n_{th}}$, with $N=\sinh^2r+N_a(\cosh^2r+\sinh^2r)$ and $M=e^{i\theta}(1+2N_a)\sinh r\cosh r$. Owing to the linear Gaussian nature of the system, its steady state is fully characterized by the covariance matrix $V$, which satisfies $AV+VA^{T}=-D$ \cite{RevModPhys.86.1391,Serafini2023}. Here, the drift matrix $A$ incorporates both the cavity-mediated coherent and dissipative couplings, while the diffusion matrix $D$ accounts for the correlated diffusion induced by the squeezed input field. Detailed analytical derivations are provided in the Supplemental Material.

Without loss of generality, we consider an antisymmetric detuning configuration, $\Delta_{m_1}=\delta_m/2$ and $\Delta_{m_2}=-\delta_m/2$, such that the two magnon modes have a relative detuning $\delta_m$. Solving the steady-state Lyapunov equation without imposing any prior assumption on the relative magnitudes of the cavity decay rate $\kappa$ and detuning $\Delta_c$, the quantum synchronization measure can be obtained in the exact form
\begin{widetext}
\begin{equation}
S_q
=
\frac{
\gamma(\gamma+4\Gamma_c)\big[(\gamma+2\Gamma_c)^2+4\Omega_c^2\big]
+\delta_m^2(\gamma+2\Gamma_c)^2
}{
\gamma T_m
\Big\{
(\gamma+4\Gamma_c)\big[(\gamma+2\Gamma_c)^2+4\Omega_c^2\big]
+\delta_m^2(\gamma+2\Gamma_c)
\Big\}
+2\delta_m^2\Gamma_c(\gamma+2\Gamma_c)T_c
}.
\label{eq:Sq_compact}
\end{equation}
\end{widetext}
Here, $T_m=2n_{\mathrm{th}}+1$ and $T_c=(2N_a+1)\cosh(2r)$. This result can be recast in the unified form
\begin{equation}
S_q=\frac{\gamma \mathcal{D}_{\mathrm{eff}}+2\Gamma_c\delta_m^2}
{\gamma T_m\mathcal{D}_{\mathrm{eff}}+2\Gamma_c\delta_m^2T_c},\label{eq:Sq_compact-1}
\end{equation}
where $\mathcal{D}_{\mathrm{eff}}=\delta_m^2+\mathcal{K}_{\mathrm{eff}}$. The quantity $\mathcal{K}_{\mathrm{eff}}$ captures how effectively the cavity-mediated channel acts on the relative magnon dynamics, with its explicit form determined by the relevant parameter regime:
\begin{equation}
\mathcal{K}_{\mathrm{eff}}=
\begin{cases}
(\gamma+2\Gamma_c)(\gamma+4\Gamma_c),
& \text{bad-cavity regime},\\[4pt]
(\gamma+2\Gamma_c)(\gamma+4\Gamma_c)(1+\alpha),
& \text{competition regime},\\[4pt]
\gamma^2+4\Omega_c^2,
& \text{large-detuning regime},
\end{cases}\nonumber
\end{equation}
where $\alpha=\frac{4\Omega_c^2}{(\gamma+2\Gamma_c)^2}$. This unified expression shows that the bad-cavity and large-detuning regimes do not require separate effective theories: their steady-state quantum synchronization is captured by the same dynamical framework, with the distinction between the regimes encoded solely in the corresponding form of $\mathcal{K}_{\mathrm{eff}}$. Thus, rather than representing independent dynamical limits, the bad-cavity and large-detuning regimes constitute different parameter-scale manifestations of the same underlying synchronization mechanism.

Since the analytical expression for $S_q$ in Eq.~(\ref{eq:Sq_compact}) relies on adiabatic elimination, we first quantify the accuracy of this reduction. We obtain the steady-state covariance matrices from the full $6\times6$ model and the reduced $4\times4$ model, and define their difference in the synchronization measure as $\Delta S_q \equiv \left| S_q^{(6\times6)}-S_q^{(4\times4)} \right|$. Figures~\ref{fig:adiabatic_validation}(a) and (b) show $S_q$ as functions of $\kappa$ at fixed $\Delta_c$ and of $\Delta_c$ at fixed $\kappa$, respectively. The two calculations are nearly indistinguishable over the parameter ranges of interest. The corresponding deviation in the $(\kappa,\Delta_c)$ plane is shown in Fig.~\ref{fig:adiabatic_validation}(c). It rapidly decreases with increasing separation between the cavity and magnon dynamical scales and remains negligible for $\lambda_{\mathrm{ad}}\gtrsim10$. We therefore use the reduced description throughout the remainder of the paper, with $\lambda_{\mathrm{ad}}\gtrsim10$ adopted as a conservative criterion for its validity. The contours at $\lambda_{\mathrm{ad}}=5$ and $10$ merely indicate the degree of timescale separation and are not intended as sharp boundaries of the adiabatic regime.

\textit{Bad-cavity limit: dissipation-dominated.—} In the bad-cavity limit, \(\kappa\gg|\Delta_c|, g\), the cavity-mediated couplings reduce to
$\Gamma_c\simeq\frac{2g^2}{\kappa},
\Omega_c\simeq\frac{4g^2\Delta_c}{\kappa^2},
\left|\frac{\Omega_c}{\Gamma_c}\right|
\simeq\frac{2|\Delta_c|}{\kappa}\ll1.
$ Coherent exchange is therefore negligible, and the dynamics are governed predominantly by collective dissipation, yielding the exact analytical expression for \(S_q\) in Eq.~\eqref{eq:Sq_compact-1}. The underlying mechanism becomes transparent in the bright--dark basis, \(\hat m_\pm=(\hat m_1\pm\hat m_2)/\sqrt2\), in which the collective dissipative channel couples exclusively to the bright mode \(\hat m_+\), while the detuning-induced term $H_\delta=\frac{\delta_m}{2}(\hat m_+^\dagger\hat m_-+\hat m_-^\dagger\hat m_+)$ hybridizes the bright and dark modes. Thus, for \(\delta_m=0\), the relative mode \(\hat m_-\) forms a dark mode and is completely decoupled from the collective reservoir. A finite detuning \(\delta_m\) breaks this decoupling, allowing the shared dissipative channel and its associated input fluctuations to act on the relative mode and thereby determine the synchronization dynamics.

\begin{figure}[t]
\centering
\includegraphics[width=\columnwidth]{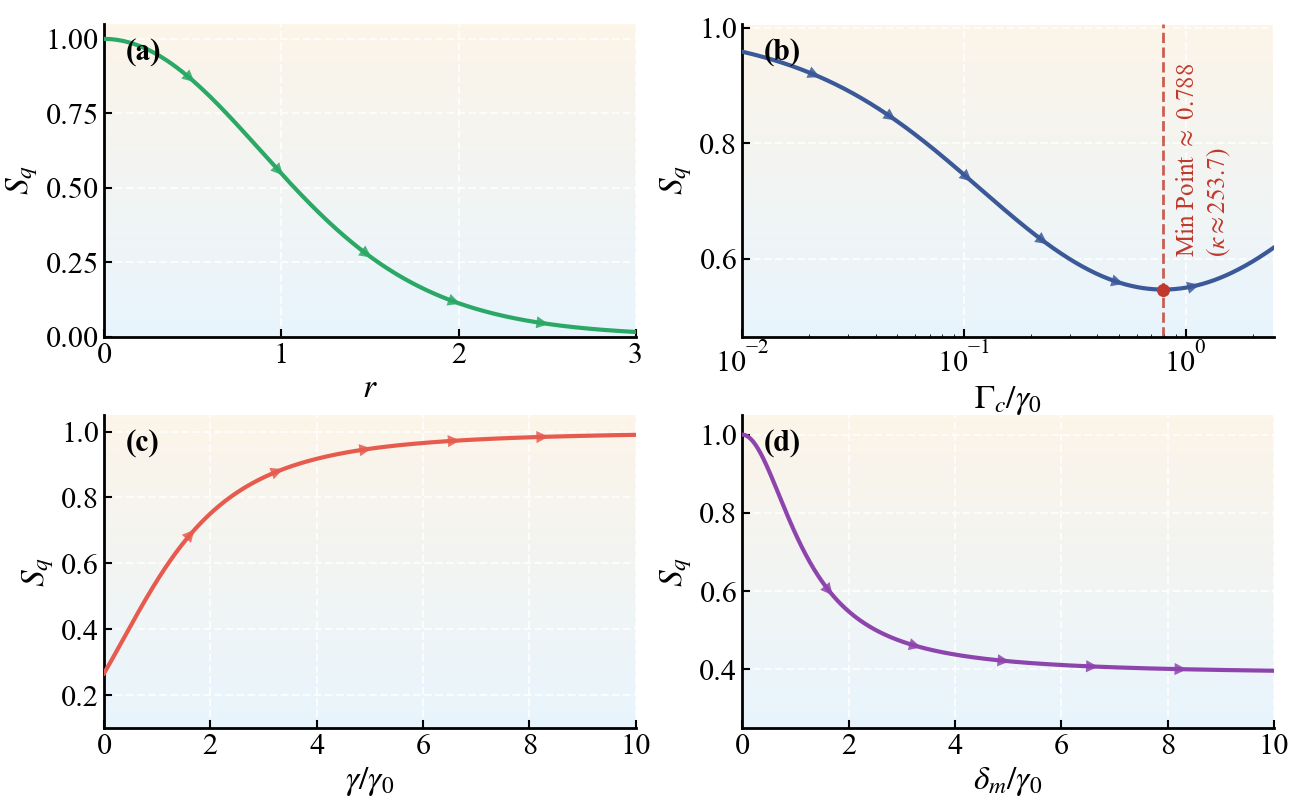}
\caption{
Dependence of the analytical synchronization measure $S_q$ on the relevant parameters in the bad-cavity regime.
The reference parameters are $\delta_m=2.0\gamma_0$, $\Delta_c=0$, and
$\kappa=300\gamma_0$.
(a) $S_q$ as a function of $r$;
(b) $S_q$ as a function of $\Gamma_c$, obtained by varying $\kappa$, with the red circle marking the analytical extremum
$\Gamma_c^*\simeq0.791\gamma_0$;
(c) $S_q$ as a function of $\gamma$;
(d) $S_q$ as a function of $\delta_m$.
}\label{fig:Sq_badcavity_4panels}
\end{figure}

The analytical expression reveals a clear interplay between dissipation, noise, and synchronization. For \(\delta_m\neq0\), increasing the squeezing strength enhances relative-mode fluctuations and suppresses \(S_q\), with \(S_q\propto e^{-2r}\) in the strong-squeezing limit \(r\gg1\). When \(\delta_m=0\), the relative mode becomes dark and remains unaffected by the shared reservoir, yielding \(S_q=1/(1+2n_{\rm th})\), which reaches unity at zero temperature. The dependence on the collective dissipation rate \(\Gamma_c\) is nonmonotonic, with an extremum at $
\Gamma_c^*=\frac{\sqrt{\delta_m^2+\gamma^2}}{2\sqrt2},$
reflecting the competition between suppression of relative fluctuations and the concomitant transfer of input noise through the shared reservoir. In contrast, local dissipation monotonically increases \(S_q\), with $
\lim_{\gamma\to\infty}S_q=\frac{1}{1+2n_{\rm th}}.$
Notably, \(S_q\to1\) as \(\kappa\to\infty\), since \(\Gamma_c,\Omega_c\to0\) and the cavity-mediated correlations disappear. This unity limit therefore signifies decoupling, rather than arbitrarily strong synchronization. The analytical predictions are in excellent agreement with the numerical results in Fig.~\ref{fig:Sq_badcavity_4panels}. Unless otherwise specified, we set \(\gamma_0/2\pi=1~\mathrm{MHz}\), \(\delta_m=6\gamma_0\), \(g=10\gamma_0\), \(\gamma=\gamma_0\), \(r=1\), \(N_a=0\), \(\omega_s/2\pi=10~\mathrm{GHz}\), and \(T=5~\mathrm{mK}\).

\textit{Mixed regime: interplay of collective dissipation and coherent exchange.—} The relative weight of collective dissipation and coherent exchange is fixed by $\frac{\Gamma_c}{|\Omega_c|}=\frac{\kappa}{2|\Delta_c|}.$ When the two rates are comparable, neither channel can be neglected and the system enters a mixed regime. Expanding to first order in $\alpha=\frac{4\Omega_c^2}{(\gamma+2\Gamma_c)^2}$, the synchronization measure reduces to Eq.~\eqref{eq:Sq_compact-1}. Here, \(\alpha\) characterizes the coherent correction to the relative-mode dynamics rather than a simple renormalization of the dissipative rate. Finite cavity detuning simultaneously generates coherent exchange and modifies the reservoir response, while collective dissipation remains active in stabilizing the relative mode.

\begin{figure}[t]
\centering
\includegraphics[width=\columnwidth]{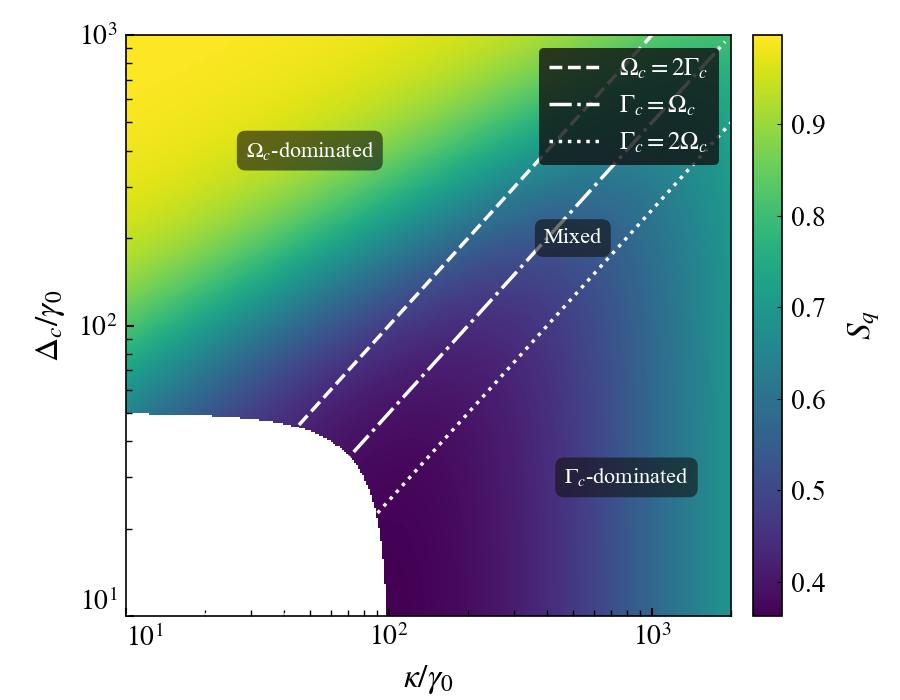}
\caption{
Steady-state synchronization measure $S_q$ obtained from the effective model after adiabatic elimination, shown in the $(\kappa,\Delta_c)$ parameter space.
The white region corresponds to $\lambda_{\mathrm{ad}} =\sqrt{\kappa^2/4+\Delta_c^2}/g<5$ and is excluded from the physical interpretation of the reduced model according to the adiabatic criterion adopted here.
To identify the parameter range in which collective dissipation and coherent exchange have comparable strengths, we define
$\frac{1}{2}\lesssim\Gamma_c/|\Omega_c|\lesssim2$
as the mixed regime.
The three reference trajectories correspond to
$\kappa=|\Delta_c|$,
$\kappa=2|\Delta_c|$,
and
$\kappa=4|\Delta_c|$,
yielding
$|\Omega_c|=2\Gamma_c$,
$|\Omega_c|=\Gamma_c$,
and
$\Gamma_c=2|\Omega_c|$,
respectively.
The two outer trajectories mark the representative boundaries of the mixed regime, while the central trajectory identifies the point at which the two effective processes have equal strengths.
The color scale represents the steady-state synchronization measure $S_q$.
} \label{fig:Sq_kappa_Deltac_map}
\end{figure}

The full analytical result in Eq.~\eqref{eq:Sq_compact} is plotted over the \((\kappa,\Delta_c)\) plane in Fig.~\ref{fig:Sq_kappa_Deltac_map}. We define$\frac{1}{2}\lesssim\frac{\Gamma_c}{|\Omega_c|}\lesssim2$
as a representative mixed region, centered at \(\kappa=2|\Delta_c|\), where \(\Gamma_c=|\Omega_c|\). This criterion specifies comparable channel weights rather than a dynamical phase boundary. Importantly, the synchronization response within this region is determined by the absolute strengths of the two channels and by the accompanying noise transfer, so that identical values of \(\Gamma_c/|\Omega_c|\) need not produce identical synchronization. This distinction is evident from the two parameter scans in Fig.~\ref{fig:adiabatic_validation}. At fixed \(\Delta_c\), increasing \(\kappa\) enhances \(\Gamma_c\) up to \(\kappa=2|\Delta_c|\), where both the dissipative coupling and the transferred noise, \(|\eta|^2=2\Gamma_c\), are maximal. Their competition produces a nonmonotonic synchronization response around the mixed region. For still larger \(\kappa\), both cavity-mediated couplings weaken and \(S_q\) increases toward the decoupled limit. At fixed \(\kappa\), by contrast, increasing \(|\Delta_c|\) suppresses \(\Gamma_c\) and its associated noise transfer while \(|\Omega_c|\) peaks at \(|\Delta_c|=\kappa/2\), driving a continuous crossover toward coherent-exchange-dominated dynamics. The mixed regime is therefore not determined by a single dimensionless ratio. Rather, synchronization reflects the joint balance of coherent exchange, collective dissipation, and noise transfer, with its response depending explicitly on the trajectory through parameter space.

\textit{Large-detuning limit: coherent-exchange dominated.—} In the opposite limit, $ |\Delta_c|\gg\kappa,g $, the cavity-mediated couplings become $\Omega_c\simeq\frac{g^2}{\Delta_c},
\Gamma_c\simeq\frac{g^2\kappa}{2\Delta_c^2}$, so that $\frac{\Gamma_c}{|\Omega_c|}
\simeq\frac{\kappa}{2|\Delta_c|}
\ll1$. The dynamics are therefore coherent-exchange dominated in terms of the relative channel strengths. Notably, however, $|\Omega_c|\propto|\Delta_c|^{-1} $ decreases with detuning, whereas both the collective dissipation and cavity-noise transfer, $\Gamma_c,|\eta|^2\propto|\Delta_c|^{-2}$ , decay more rapidly. Thus, large detuning does not enhance the coherent interaction; rather, it preferentially suppresses the cavity-induced noise that contaminates the relative mode. The synchronization measure is then given by Eq.~\eqref{eq:Sq_compact-1}, which, for a pure squeezed input at low temperature, yields $1-S_q \simeq
\frac{g^2\kappa}{\Delta_c^2}
\frac{\delta_m^2}{\gamma(\gamma^2+\delta_m^2)}
\left[\cosh(2r)-1\right]$.
The leading correction is therefore entirely controlled by the residual transfer of cavity fluctuations into the relative mode. Because $|\eta|^2=2\Gamma_c$, this noise contribution is rapidly suppressed with increasing detuning, allowing $S_q$ to approach unity even though the coherent exchange itself becomes weaker.

At finite detuning, this asymptotic behavior is modified by the simultaneous variation of the two mediated channels. Increasing $ |\Omega_c| $ requires reducing $|\Delta_c|$, which enhances both $\Gamma_c$ and $ |\eta|^2 $. The resulting competition between coherent control of the relative dynamics and cavity-noise injection produces a nonmonotonic dependence of $S_q$ on $|\Omega_c| $. The leading $O(\Delta_c^{-2})$ expansion captures the asymptotic scaling but not the finite-detuning minimum, which arises from higher-order corrections to the full steady-state solution. This minimum is therefore a finite-detuning consequence of coherent--dissipative competition rather than a universal feature of the strict large-detuning limit.

\begin{figure}
\centering
\includegraphics[width=\linewidth]{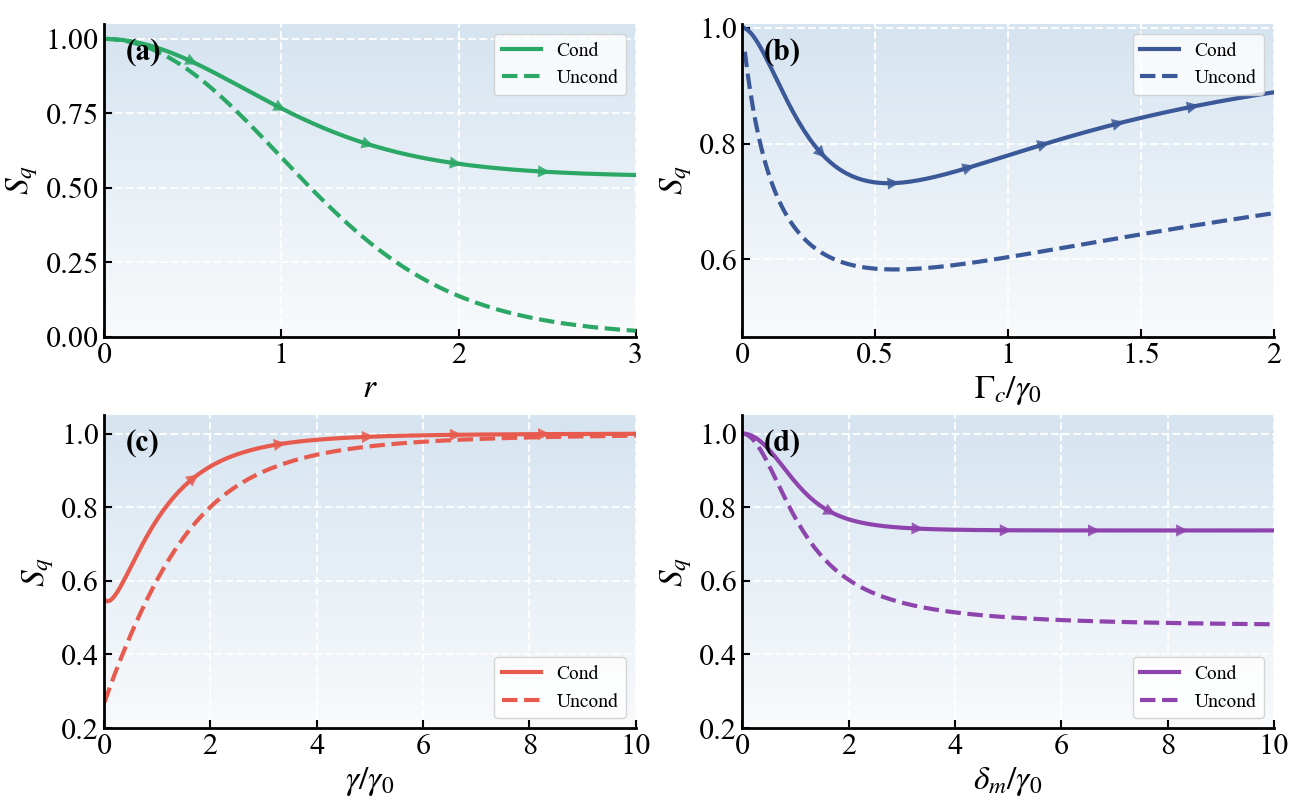}
\caption{
Enhancement of quantum synchronization by continuous heterodyne detection.
Solid and dashed curves denote the conditional and unconditional results, respectively.
Parameters are $\kappa=300\gamma_0$, $\Delta_c=150\gamma_0$, $\delta_m=2\gamma_0$, and $\zeta=1$ (unit detection efficiency).
The four panels show the synchronization measure as a function of
(a) $r$,
(b) $\Gamma_c$ varied by tuning $g$,
(c) $\gamma$,
and (d) $\delta_m$.
}
\label{fig:conditional_sync}
\end{figure}

\begin{figure}[t]
    \centering
    \includegraphics[width=0.85\linewidth]{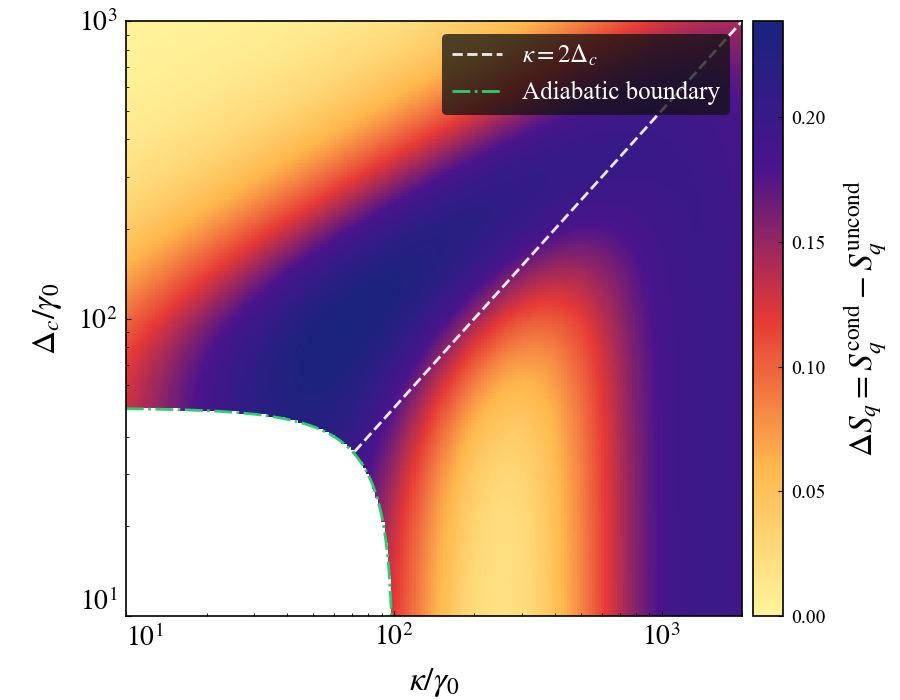}
    \caption{
    Distribution of the measurement-induced enhancement of quantum synchronization in the $(\kappa,\Delta_c)$ parameter plane.
    The color scale represents the measurement-induced synchronization gain
    $\Delta S_q=S_q^{\mathrm{cond}}-S_q^{\mathrm{uncond}}$.
    The dashed line $\kappa=2\Delta_c$ corresponds to the equal-weight trajectory
    $\Gamma_c=|\Omega_c|$.
    The white region denotes the parameter range with
    $\lambda_{\mathrm{ad}}<5$, where the adiabatic-elimination condition is not sufficiently satisfied and the reduced model is therefore not used for quantitative interpretation.
    The remaining parameters are $\delta_m=2\gamma_0$ and
    $\zeta=1$.
    }
\label{fig:delta_Sq_map}
\end{figure}

\textit{Continuous measurement enhanced quantum synchronization.—} The synchronization measure considered above is based on the unconditional steady-state covariance matrix and therefore describes an ensemble-averaged state in which information carried by the environmental output is discarded. Here, the two magnons couple to a common dissipative channel mediated by the cavity, which acts simultaneously as a source of correlated fluctuations and as an information carrier of the collective magnon dynamics. Continuous monitoring of the output field can therefore recover information lost in the unconditional description and condition the magnon state on the measurement record. We implement this procedure by continuously heterodyne detecting the cavity output field. The resulting conditional dynamics can be described by a measurement-conditioned master equation, or equivalently by the quantum-trajectory formalism \cite{wiseman2009quantum,c5wj-4s73,PhysRevE.108.024204,Kato2021}. For the linear Gaussian system considered here, the conditional covariance matrix can be obtained analytically \cite{wiseman2009quantum,Genoni2016}, with details given in the Supplemental Material.

\textcolor{black}{As shown in Fig.~\ref{fig:conditional_sync}, continuous heterodyne detection enhances the quantum synchronization measure, with
$S_q^{\mathrm{cond}} > S_q^{\mathrm{uncond}}$
throughout the parameter range considered. We focus on the regime
$\Gamma_c = |\Omega_c|$,
where coherent exchange and collective dissipation have comparable strengths, as a representative setting to examine the measurement-induced enhancement.}
 The measurement record provides information that is inaccessible in the unconditional state and reduces the uncertainty of the relative mode through conditional state updating. This information becomes particularly relevant for finite \(\delta_m\), which hybridizes the bright and dark modes and thereby transfers relative-mode information to the monitored output.

We quantify the resulting gain by $\Delta S_q=S_q^{\mathrm{cond}}-S_q^{\mathrm{uncond}}.$ As shown in Fig.~\ref{fig:delta_Sq_map}, \(\Delta S_q\) is strongly controlled by the cavity parameters. In the large-detuning regime, \(\Gamma_c\) and the corresponding information flow to the output both diminish, leaving little information available for conditioning. In the bad-cavity limit, $
\Gamma_c\simeq\frac{2g^2}{\kappa},\Omega_c\simeq\frac{4g^2\Delta_c}{\kappa^2},$
the coherent exchange is suppressed more rapidly than the collective dissipation as \(\kappa\) increases. The unconditional dynamics are consequently dominated by the shared dissipative fluctuations, while the output channel retains appreciable information about the magnon motion. Continuous monitoring can then recover this information and substantially reduce the synchronization error. Thus, the same dissipative channel that limits unconditional synchronization can, when monitored, become a resource for enhancing it.

\textit{Conclusions—}We have established a general framework for quantum synchronization mediated by a fast common reservoir. Adiabatic elimination separates the resulting coherent exchange and collective dissipation, revealing a continuous crossover between two distinct synchronization mechanisms. Synchronization is determined by their combined strengths and the associated noise transfer; notably, large detuning can enhance synchronization by suppressing reservoir fluctuations faster than coherent exchange. Continuous monitoring further exploits the dual noise--information role of the reservoir, converting environmental information into enhanced conditional synchronization.\textcolor{black}{The magnitude of this enhancement exhibits a pronounced parameter dependence, reflecting the interplay among coherent exchange, collective dissipation, noise transfer, and measurement-induced information gain.}with the largest gain near the coherent--dissipative crossover. These results provide a general route to understanding and controlling synchronization in open quantum systems with shared reservoirs.

 \bibliography{references}

\end{document}